# On Practical Aspects of Mobile Data Offloading to Wi-Fi Networks

Adnan Aijaz[†], Nazir Uddin[‡], Oliver Holland[†], and A. Hamid Aghvami[†]

[†]Centre for Telecommunications Research, King's College London, London WC2R 2LS, UK
[‡]Nokia Siemens Networks, Indonesia
{adnan.aijaz, oliver.holland, hamid.aghvami}@kcl.ac.uk, nazir.din@nsn.com

*Abstract*

Data traffic over cellular networks is exhibiting an ongoing exponential growth, increasing by an order of magnitude every year and has already surpassed voice traffic. This increase in data traffic demand has led to a need for solutions to enhance capacity provision, whereby traffic offloading to Wi-Fi is one means that can enhance realised capacity. Though offloading to Wi-Fi networks has matured over the years, a number of challenges are still being faced by operators to its realization. In this article, we carry out a survey of the practical challenges faced by operators in data traffic offloading to Wi-Fi networks. We also provide recommendations to successfully address these challenges.

## I. Introduction

In recent years, data traffic transmitted over cellular/mobile networks has seen a continuous exponential growth increasing by an order of magnitude every year. According to Cisco forecasts [1], global mobile data traffic is expected to grow to 15.9 exabytes (1 exa = $10^{18}$) per month by 2018, which is an 11-fold increase over 2013. This unprecedented growth of data traffic can be attributed to a number of factors. A first factor is the introduction of high-end devices such as smartphones, tablets, laptops, handheld gaming consoles, etc. that can multiply traffic (e.g., a tablet can generate up to 120 times the traffic generated by a basic feature phone). Secondly, the growth in mobile network connection speeds that increase the average traffic per device (e.g., in 2013, a 4G connection generated 14.5 times more traffic than a non-4G connection). Thirdly, the rise of mobile video content which has higher bit rates than other mobile content types. Mobile video traffic has already surpassed 50% of total

2mobile data traffic and continues to increase due to a number of technological advancements including the larger screen sizes of smartphones as well as the optimization of mobile video content for mobile devices that enhance a user's viewing experience. Fourth is the availability of mobile broadband services at prices and speeds comparable to those of fixed broadband, together with the increasing trend towards ubiquitous mobility. Last, but not the least, the widespread adoption of Machine-to-Machine (M2M) [2] technologies across a range of industries is another contributing factor.

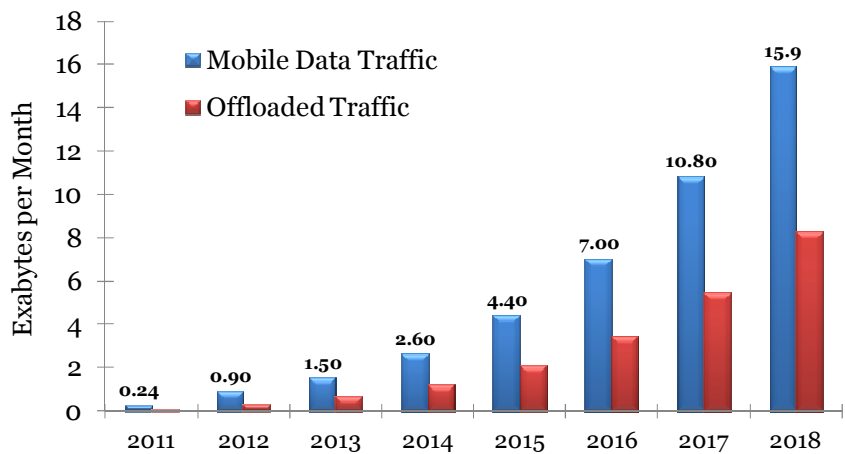

Fig 1: Global mobile data traffic along with current and predicted offloaded traffic [1]

Mobile data offloading or simply data offloading refers to the use of complementary network technologies and innovative techniques for delivery of data originally targeted for mobile/cellular networks in order to alleviate congestion and making better use of available network resources. The objective is to maintain Quality-of-Service (QoS) for customers, while also reducing the cost and impact of carrying capacity hungry services on the mobile network. The traditional approach of scaling network capacity with additional network equipment is always available, but not cost effective and viable considering the pace at which the demand for data services is increasing. It is expected that mobile data offloading will become a key industry segment in near future as the data traffic on mobile networks continues to increase rapidly.

Most mobile operators worldwide have already started to implement an offloading solution or strategy. In our previous work [3], we carried out a comprehensive survey of the state of the art in mobile data offloading, covering both technological and business aspects. Key technologies include Wi-Fi, femtocells, and IP flow mobility. As the growth of data traffic also creates challenges for the backhaul of cellular networks therefore techniques such as core network offloading and media optimization are also gaining popularity.

Offloading to Wi-Fi networks has evolved as one of the mature data offloading solutions, besides femtocells. In 2013, 45% of global mobile data traffic was offloaded onto fixed networks through Wi-Fi or small cells and the offloaded volume is expected to grow to 52% by 2018 [1]. Wi-Fi comes as a natural solution for offloading due to built-in Wi-Fi capabilities of smartphones. Due to degradation of cellular services in overloaded areas, an increasing number of users are already using Wi-Fi to access Internet services for better experience. From service provider's perspective, Wi-Fi is attractive because it allows data traffic to be shifted from expensive licensed bands to free unlicensed bands (2.4GHz and 5GHz). Studies have shown that expanding network using Wi-Fi is significantly less expensive compared to a network build-out. Nowadays, Wi-Fi is undergoing a paradigm shift towards ubiquity and outdoor/city-wide Wi-Fi networks are gaining popularity. An increasing number of mobile operators have started deploying a Wi-Fi offloading solution. The most popular strategy is the extension of an operator's access network to include hotspots directly managed by the operator. Our objective in this article is to identify the practical challenges faced by operators while deploying a Wi-Fi offloading solution. We begin our discussion by explaining different approaches of offloading traffic to Wi-Fi networks. After that we outline the main challenges from network, device, and regulatory perspectives. We also provide recommendations for effectively addressing these challenges. Finally we conclude the article.

## II. MOBILE DATA OFFLOADING VIA WI-FI NETWORKS

### A. Wi-Fi Offloading Approaches

There are three main approaches for operators to offload data traffic onto Wi-Fi networks, depending upon the level of integration between Wi-Fi and cellular networks [3]. The first approach is the *network bypass* or the *unmanaged data offloading* in which case the users' data is transparently moved onto the Wi-Fi network, whenever they are in Wi-Fi coverage, completely bypassing the (cellular) core network for data services. Voice services; on the other hand continue to be delivered via the cellular core network. Whilst this approach seems attractive as it does not require the deployment of any network equipment, it has some drawbacks. Firstly, the operator loses visibility (and hence control) of its subscribers whenever they are on the Wi-Fi network. Secondly, the operator is unable to deliver any subscribed content (Blackberry, corporate VPN, ringtones etc) leading to potential loss of revenue. Despite its disadvantages, this approach can be adopted as an immediate offloading solution due to its ease of deployment. It is also attractive from users' perspective due to control over data connectivity. It should be noted that this approach is similar to the users switching on the Wi-Fi interface, whenever they are in Wi-Fi coverage for better experience. The operator can deploy such an offloading solution by simply placing an application in handsets that switches on the Wi-Fi interface on detecting Wi-Fi coverage.

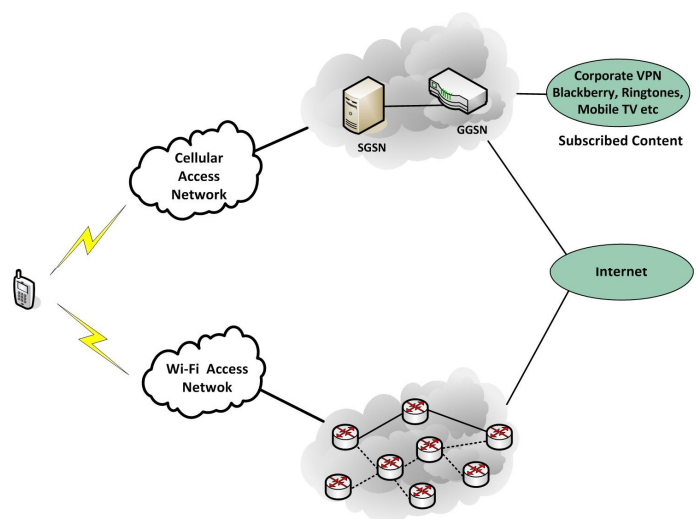

(a)

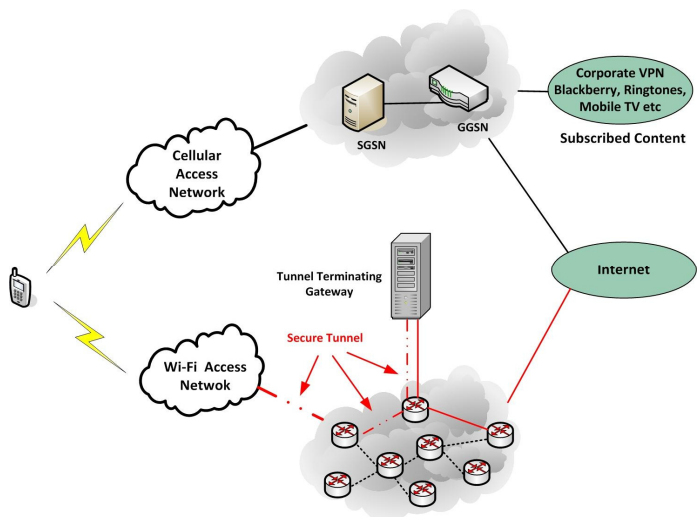

(b)

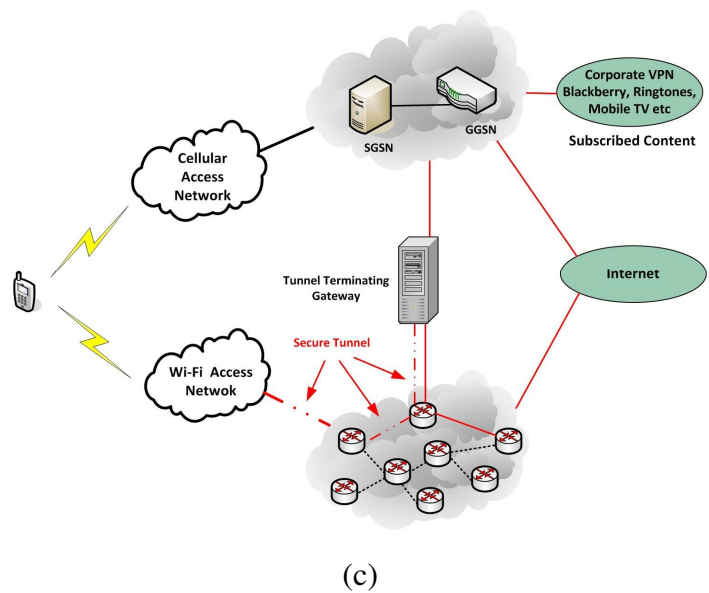

(c)

Fig 2: Mobile data offloading via Wi-Fi, (a) network bypass, (b) managed approach, (c) integrated approach

A *managed data offloading* approach can be adopted by those operators who do not want to lose control and visibility of their subscribers. This could be due to a number of reasons. Some operators provide metered network access which requires subscriber control. Others deliver services such as parental control/filtering, that cannot be provided in a completely desegregated network. Some operators simply want to be aware of subscribers browsing habits for targeted marketing reasons. All this can be achieved by placing an intelligent session aware gateway through which the subscriber's Wi-Fi session traverses on its way to the Internet. Complete integration of cellular and Wi-Fi networks is not required in this case. While the operator gains control of subscribers, it still cannot deliver any subscribed content.

An *integrated data offload* approach provides the operator with full control over subscribers as well as the ability to deliver any subscribed content while the users are on the Wi-Fi network. This is achieved by the integration of cellular and Wi-Fi networks so that a bridge can be formed between the two networks through which data flow can be established. There are two architectures for coupling cellular and Wi-Fi networks; loose coupling and tight coupling. In loose coupling architecture, the networks are independent requiring no major cooperation between them. The Wi-Fi network is connected indirectly to the cellular core network through an external IP network such as the Internet. Service connectivity is provided by roaming between the two networks. On the other hand, in a tightly coupled system, the networks share a common core and majority of network functions such as vertical handover, resource management, and billing are controlled and managed centrally.

The 3GPP I-WLAN [1]standard [4] defines the basic principles for managing Wi-Fi networks and accessing operator services in an integrated (tightly coupled) cellular and Wi-Fi system. The underlying concept is to establish a controlled tunnel between the mobile device and the infrastructure of mobile network operator as shown in Fig. 2c. This is further elaborated in Fig. 3 where we consider the Evolved Packet Core (EPC) of the LTE network. Since Wi-Fi hotspots are generally not considered secure, the I-WLAN standard specifies how a secure Virtual Private Network (VPN) tunnel can be established between an authorized mobile device (connected to Wi-Fi hot spot) and a VPN gateway residing in the EPC. This VPN gateway is known as the evolved Packet Data Gateway (ePDG). The traffic is tunnelled between the operator services via a WLAN Access Gateway (WAG) and the ePDG. The WAG acts as a dynamically configured firewall, while the ePDG as a tunnelling end point.

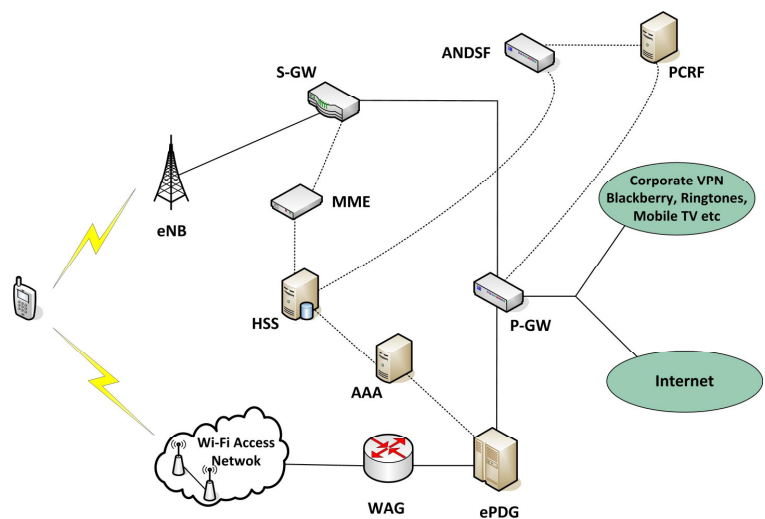

Fig 3: 3GPP I-WLAN specified tightly coupled cellular (LTE) – Wi-Fi system

B. *Enhancing Wi-Fi for Mobile Data Offloading*

Recently, 3GPP developed the Access Network Discovery and Selection Function (ANDSF) [10] as part of the EPC for cellular networks. ANDSF offers a dynamic offloading capability by enabling mobile devices to discover and connect to non-3GPP networks (e.g., Wi-Fi and WiMAX). It provides network discovery information of access networks that may be present in the vicinity of the mobile device along with the network selection rules. The 3GPPP ANDSF Management Object (MO) specifies the format of information exchange between the ANDSF server and the mobile device. Moreover, ANDSF also provides inter-system mobility (for devices that do not support more than one network interface at a time)

---

[1] To the best of our knowledge, no operator has implemented the 3GPP I-WLAN standard yet

and inter-system routing (devices capable of connecting to multiple access networks at a time) policies.

Another notable development in context of Wi-Fi offloading is the IEEE 802.11u standard [11] that provides capability to mobile devices for selecting an appropriate Wi-Fi access point before actually associating with it. This is particularly useful in locations where several Wi-Fi access points exist and the mobile device has no way of determining the best access point as per its requirements. At present, Wi-Fi access points only broadcast their SSIDs in the beacon. However, 802.11u allows a much richer set of capabilities to be advertised by Wi-Fi access points. This includes information regarding the type of network (private, public, or paid public, etc.), information about the current load (which can give a hint about the level of congestion on the access point), or details about the roaming agreements with other service providers etc. Moreover, 802.11u also introduces Access Network Query Protocol (ANQP), which allows a mobile device to run a query/response session with the access point before associating with it.

### III. PRACTICAL CHALLENGES FACED BY OPERATORS IN WI-FI OFFLOADING

#### A. When to Introduce Offloading?

Cellular operators (2G and 3G alike) are receiving offers from Wi-Fi equipment vendors on different choices of using Wi-Fi as a medium to offload data traffic in congested data hotspots. These hotspots are not limited to indoor environment like shopping malls, airports, and train stations etc. but also include outdoor areas like financial districts and stadiums. However one of the key challenges operators face today is when is the right time to deploy Wi-Fi offloading in these areas. The questions surrounding them are:

- At what loading of the cellular network should offloading to Wi-Fi be initiated? Do they let the base station reach maximum capacity for data?
- Do they look at cellular spectrum congestion/interference?
- Do they decide based on expansion cost of base station transceivers and its associated service charges?
- Can the decision be driven by the potential for Wi-Fi offloading to reduce other forms of CapEx, such as backhaul provision?

The reasons are not purely technical or commercial, but a mix of both. The frequency spectrum is already reaching its limits in terms of bits/s/Hz. With addition of more access network capacity, backhauling needs continue to rise. Moreover, the distribution of traffic load varies both spatially and temporally and may not always be well suited to the use of fixed, limited coverage and limited spectral capacity Wi-Fi solutions as a reliable backup. Apart from this, groups of users are spread over different areas. Therefore, setting a threshold for traffic (e.g., in terms of Mbps per base station) can become a challenge if the base station stands isolated amongst its neighbours.

All the above factors contribute to increase in cost ($ per bit) for data and with falling revenue per bit from data traffic, operators are faced with a dilemma of deciding when to introduce Wi-Fi as a cost effective means of carrying data traffic.

### B. Where to Offload?

Based on the current traffic per base station and total traffic for a cluster (of base stations), operators make a decision looking at the forecast of number and types of subscribers within that area. Selection of area is easier if the traffic is purely indoor. However, if the traffic is hybrid or purely outdoor then a further assessment needs to be done to pinpoint specific potential high data traffic areas under the base station. Taking into account the coverage potential of access points within the regulatory constraints of transmit power, it is important to identify ideal areas. As a practice, operators tend to evaluate if additional outdoor capacity can be handled by minimal re-design effort on cellular networks. If that is not achievable then it is a candidate for Wi-Fi base station.

### C. Availability and Limitations of Wi-Fi Planning Tools

Once the potential areas have been identified through physical survey based on the data traffic distribution from the cellular network, there is a need to plan the deployment of Wi-Fi access points. Such planning activities require the use of planning tools that can predict coverage based on the demographics. Currently there is lack of planning tools available to design Wi-Fi networks for an outdoor environment. Most available tools are optimized for indoor environment with specific 2-dimensional views of the building floors and their types. Indoor environment are far more controlled and predictable compared to the outdoor. Indoor prediction tools use set of variables on types of indoor RF environment with glass partitions, wooden separations, cement, or concrete walls. Indoor planning tools by *AirMagnet / Fluke Networks* are widely used by both vendors and operators for the purpose of indoor RF prediction and coverage analysis. *iBwave* is also polishing its design and propagation suite to generate coverage predictions. However, these tools or similar tools in the market are built, designed, and optimized for indoor only. This means the prediction models are meant for simple RF environment with smaller number of variables representing environment. They rely on floor plans to simulate RF environment and have limitations on calculating interference i.e., only non-Wi-Fi devices are considered. Another limitation is that they are not designed to provide any dimensioning numbers that are used for capacity planning.

Outdoor prediction brings a new set of challenges with far more complex RF environment that current Wi-Fi RF propagation tools cannot handle. Some vendors have modified their indoor planning tools for outdoor coverage calculations but at best represent an interim way of predicting coverage. However these are better than trial and error methods which are based solely on placing an access point at a central point and trying to cover as much area as possible. Market is thus ripe for a planning tool that can address the needs of data offloading on hotspots not just by planning access points but engaging overlay layers of 3G/LTE to recommend possible candidate locations. Another bottleneck for Wi-Fi in terms of planning tools is the investment in the tools itself. Earlier belief being that an unlicensed band does not warrant investment on design tools is slowly fading away but that is going to take some time. Tier-1[2] operators are taking lead on evaluating possible upgrades or add-on modules to their existing portfolio of planning platforms. Other players in the value chain like system integrators, equipment vendors, etc. are trying to forge partnership and relationships to acquire this competence and are at nascence.

### D. Backhauling Challenges

Offloading traffic on Wi-Fi also presents challenges for backhauling as briefly highlighted above. Using licensed means to backhaul unlicensed Wi-Fi traffic is an interesting question for operators. However for most operators answer depends on the location and volume of traffic. In cases where deployment of licensed spectrum radios, or fiber, or xDSL lines are cost prohibitive or impossible to deploy due to any number of reasons, it is viable to look at unlicensed bands. Within unlicensed bands, a combination of point-to-point mesh in 5GHz and self-backhauling within 2.4GHz are popular choices.

### E. Site(s) Availability and Acquisition

Site acquisition for both indoor and outdoor Wi-Fi deployments is a challenge due to real estate permissions, diverse installation practices and guidelines, site accessibility, and costing models. Markets are not very well regulated in terms of permissions for installation and costing for sites. In most cases it is up to the building owners to decide. There are different sets of challenges for public and private real estates. Coordinating with private building owners adds another challenge where talking to a large number of building owners takes up significant time and resource to ensure reasonable terms and conditions are agreed. These terms and conditions govern the location of installation, site accessibility times during the day or night, access to power, and route of cabling.

---

[2] Tier -1 operators are group companies with presence in different parts of the world. They represent leading market share in their local markets and use latest cellular technologies in their network.

*F. Deployment Issues*

There are a number of challenges faced during the deployment of Wi-Fi access points. For indoor environments, foremost consideration is the installation of access points that are exposed on the ceiling. In case they are to be installed behind a false ceiling then the material of false ceiling is an important consideration keeping in view the signal attenuation. For aesthetic reasons its becoming common to hide access points behind a ceiling. Others factors during indoor installation like location of access panel, presence of other utilities, routing of the cable, and access to power source are important issues.

For the outdoor deployment major concerns are availability of infrastructure to place the access point and the availability of power. Cost of developing or leasing infrastructure to mount access point drives the cost of project and also impacts the backhaul choice. For street level coverage, using cellular base station sites for installing access points is a challenge. Base station sites are normally high rise and 30 – 45 meters above the ground. The transmit power of access points is not high enough to achieve a strong street level coverage. Lamp posts are a better alternative; however, the availability of lamp posts and permission to install access points poses a challenge.

*G. Regulatory Constraints*

Regulatory aspects drive the adoption of Wi-Fi as an offloading solution especially on the transmission power restrictions. Wi-Fi base station concept and other similar high transmit power options face challenges due to the limitations imposed by regulators. It also drives overall cost of the solution where far greater number of access points are needed if transmit power is curtailed to 100mW compared to 300mW [8]. Industry is also uneven in terms of regulating the operators who can deploy Wi-Fi networks. In some countries cellular operators are allowed to deploy Wi-Fi whereas in others it's only meant for Internet Service Providers (ISPs) e.g., Bangladesh. In such markets, operators are forging partnerships with ISPs to get around such a challenge. For markets with favorable regulatory policies, operators are now being assessed for the Quality-of-Service (QoS)[3] provided to end users. Certain developed markets have seen a flurry of activity where regulators are developing practices and framework to ensure a high QoS for users offloaded to Wi-Fi. These include audits and benchmarking for accessibility and retainability.

### H. Device Limitations

Market is currently proliferated by Wi-Fi devices, ranging from cellphones to laptops, desktops, cameras, tablets, TVs etc. However a large percentage of these devices are 2.4 GHz compliant. Although there has been a drive by equipment manufactures to introduce IEEE 802.11ac compliant products, it is still a long way before we can see a large percentage of devices being 5 GHz compatible. Large density of 2.4 GHz Wi-Fi devices render the effectiveness of IEEE 802.11n access points useless, since users cannot be migrated to a cleaner RF environment.

### I. One-Way Offloading

A key aspect of data offloading to Wi-Fi is the technique of migration itself. Operators are becoming acutely aware of the challenges of 'dumb' or 'manual' offloading. Manual offloading is one-way offloading where user end devices would automatically tune to Wi-Fi if they have previously used it and/or the SSID is remembered. One-way offloading has certain draw backs since it forces the users to hop on to Wi-Fi even when it does not need to. Users tend to stay connected to Wi-Fi as long as they are able to get an RF signal. This sometimes means users stay on Wi-Fi even if signal strength is below desired threshold for a reasonable data rate. Users continue to drag Wi-Fi until there is no Wi-Fi coverage.

One-way offloading can become a data black hole as it continues to attract users to switch over to Wi-Fi when QoS is worse than the cellular 3G/LTE network. It also poses challenges to operators to achieve density of Wi-Fi access point deployment, as higher number of access points would themselves create an interfering environment. What operators are very interested in looking are ways and techniques of 'intelligent' or smart handover between 3G/LTE and Wi-Fi which takes into account situation of both networks.

### J. Charging Nightmares

Charging mechanisms are an important aspect of any Wi-Fi offloading scenario. Operators are adopting a diverse set of approaches in this regard. Some operators do not intend to charge data over Wi-Fi any different than when they are on the cellular network. However for others introducing Wi-Fi as a separate layer came up with low cost solutions. In some places like airports, train stations etc. Wi-Fi is offered as a free of charge service. There is yet another group of operators who have taken middle ground where the users manually switch over to Wi-Fi and if these users belong to same cellular network they are charged at a fraction of the cost but for other users a fee is charged based on a selective package. Normally multiple SSIDs are created to separate free and premium users. Most of the 2G (GSM) operators aspiring for 3G licenses in developing markets seem to favor this approach.

---

[3] This QoS assessment is conducted by regulators through autonomous testing in terms of different KPIs.

Users which are seamlessly handing over from cellular to Wi-Fi do pose a challenge in terms of billing as they try to replicate cellular packages. Real time charging between floating users is also something that poses an accounting challenge.

*K. Wi-Fi vs DAS and Non-Wi-Fi Interference*

Recently, Distributed Antenna System (DAS) [7] has attracted a lot of attention in cellular networks. DAS refers to a network of spatially distributed antenna nodes connected to a home base station through a high bandwidth low latency dedicated connection. Intuitively, this has the effect of reducing the average distance of propagation to or from the nearest antenna, thus reducing the required uplink and downlink transmitted power for a fixed channel quality and creating more uniform coverage inside the cell.

Integrating Wi-Fi within distributed antenna system (DAS) platforms has been discussed amongst operators in order to leverage on the existing in-building installations. However the results have been discouraging because of Wi-Fi access points' receiver desensitization due to high power signal from GSM/3G.

Two main approaches have been adopted to integrate Wi-Fi within a DAS platform. For a passive DAS, integrating Wi-Fi access point's RF interface at the combiner results in capacity constraints. High power signals from GSM and 3G base stations effectively block the Wi-Fi single channel signal. Wi-Fi in such a scenario is also unable to use its MIMO capability. Further, single access point is unable to meet the capacity needs of the entire building. This approach also has not yielded any viable results even when access points were integrated with active DAS.

The second method of integration is at antenna level. In this case, one access point with RF interface for 2.4 GHz output is integrated with DAS at the antenna point. The challenges include poor Wi-Fi coverage due to high power signal from cellular services and running of long data cables to backhaul the traffic towards an access controller.

*L. Security Issues*

As the number of high-end devices (smartphones, tablets, etc.) continues to grow, the focus on security is equally important on the device, network, and the data traversing both secured and unsecured Wi-Fi networks. From an operator' perspective, carrier grade Wi-Fi requires strong security without compromising user experience. The most important issue is of user authentication which should be a seamless activity. The authentication issue becomes particularly important with roaming between Wi-Fi networks (of different service providers) as the roamed network has no access to the encryption keys used to authenticate the user.

IV. RECOMMENDATIONS FOR SUCCESSFULLY ADDRESSING THESE CHALLENGES

In order to address the challenges Wi-Fi faces to succeed as a viable alternative, the most effective way is to use a cherry picking approach. This means focusing on high data traffic areas. These areas can be then be segregated into outdoor and indoor locations to develop plans and to prioritise the deployment of access points. The roll out of sites can then take place, taking into consideration logistics, backhauling, and site priority.

The limitations of planning tools for outdoor Wi-Fi can be addressed by extensive active and passive surveying to allow for true representation of the RF environment and selection of candidates for installation near the intended hotspots. Active survey provides the projection of received quality and signal strength as it associates with the access point which is temporarily deployed at the site. These cover both 2.4 GHz and 5 GHz bands and assists to define received signal level for acceptable coverage both for small and medium screens. Use of digital maps during the planning stage enables to identify and anticipate the challenges of the site selected for access point's deployment. With tools like *Atoll* and *Mentum* integrated with digital maps predictions are more accurate to achieve good outdoor street level coverage.

Backhauling in a data dominant era has been a subject of a lot of speculation and activity. The most cost effective and proven technique to backhaul Wi-Fi access points is by using 5 GHz band of dual-band access points. Using total mesh or partial mesh helps with robustness of backhaul. On the aggregation or hub point we see licensed radios being a better solution as well as WiMAX (World-Wide Interoperability for Microwave Access) and fiber. For operators that have Wi-Fi clusters spread across a city area and possess a WiMAX band, there is a great opportunity to use it as a backhaul. With declining interest in WiMAX as an access technology, licensed frequency spectrum is available for backhauling. In some countries operators are already started using WiMAX to backhaul traffic subject to local regulations [5].

For the cases where cellular operators use their 2G/3G macro and micro base stations as co-location for Wi-Fi or where multi-mode cells are being introduced, backhaul is shared with cellular traffic. These are normally IP or hybrid microwave radios that are capable of transporting traffic. If these radios are purely TDM (Time Division Multiplexing) based, we can consider placing cell site routers (CSR) to aggregate both TDM and IP traffic. However, considering long term evolution of transport layer it is strongly recommended to use native IP radios that can reduce the overhead. We see vendors like *Cisco* and *Juniper* placing CSRs to allow using native layer 3 to transport 2G/3G/Wi-Fi traffic. An increasing number of operators are conducting trials for the use of licensed point-to-multipoint radios in 26/28 GHz bands for backhauling of multi-mode cells.

Site sharing and wholesaling had added a new dimension to Wi-Fi ecosystem where dedicated operators like *F5 Networks* and *Tikona* have found viable business models to act as wholesale operators for cellular networks and act as a Virtual Network Operator (VNO). Such scenarios warrant the use of fiber especially on aggregation nodes to ensure high availability and high capacity. This model also reduces the challenge posed by site acquisition by multiple operators as a single party acquires and maintains the sites. For outdoor Wi-Fi access point installations, operators can prioritise sites near lamp posts and also target locations where permission to install is easily available. Sectorization of access points also provide part of the solution as these can be wall mounted and positioned towards the area e.g., outdoor restaurant, bar, shops, walk ways, etc.

In an effort to move IEEE 802.11n and .11ac compliant devices to less used 5 GHz spectrum, vendors have come up with smart band steering features that push devices to a faster band. While this positively impacts the QoS, it also reduces the congestion on the 2.4 GHz spectrum. Development of devices that support 5GHz has been slow in the past but with *Apple* and *Samsung* releasing .11ac devices we can sense a pickup especially for Smart TV needs. Operators, system integrators like *NERA Telecommunication* along with their partners are coming up with innovative solutions like managing channel selection and avoiding congested channels to perform in the interference prone unlicensed band of Wi-Fi.

There are several techniques currently seen in the market to achieve intelligent handovers which are bi-directional. Simply stated, the user device will not automatically switch to Wi-Fi unless it fulfils certain quality objectives. These objectives form the basis of quality based handover rather than power based handover which are currently at offer. To complete the ecosystem we witness the emergence of connection managers for the UEs to decide on staying on 3G or switch to Wi-Fi and return back once Wi-Fi performance is degraded. *Korea Telecom* has come up with a novel approach which it exhibited during Wi-Fi Congress in South Korea recently [6].

TABLE 1: A summary of challenges and recommendations for Wi-Fi Offloading

|  | Challenges | Recommendations |
|---|---|---|
| 1. | Availability and limitations of Wi-Fi planning tools | Active and passive surveying, integration of digital maps with existing tools |
| 2. | Backhaul | WiMAX, wireless mesh networks (5GHz) |
| 3. | Site availability, acquisition, and deployment issues | Site sharing (with friendly business models), prioritizing sites near existing infrastructure, sectorization of Wi-Fi access points |
| 4. | Device limitations | Band steering solutions, introduction of 802.11ac compliant devices by vendors |
| 5. | One-way offloading | Quality-based handovers, connection managers for mobile devices, and 802.11u based Wi-Fi hotspots |
| 6. | Charging issues | Time, volume, and location based charging, incentive driven data packages, etc. |
| 7. | Authentication | EAP-SIM based authentication techniques |
| 8. | Wi-Fi and DAS Integration | Deploy a separate layer of Wi-Fi access points |

Time, volume, and location based charging methods or a combination of these packages is becoming common to encourage users to use and stay on Wi-Fi for as much of the time as possible. Operators are also testing concepts where a user within his home premises is charged less or nothing at all for Wi-Fi usage compared to when he is in the mall. Some operators use free Wi-Fi as an incentive to push users away from their congested 3G traffic. Other operators with more uniform traffic distribution keep the costs over both Wi-Fi and 3G as a same package. With the increasing focus on data usage very creative and dynamic billing packages are surfacing. New packages such as users sharing their data package within a selective social group are picking up.

Operators are making special efforts to ensure that authentication is as seamless as possible since user experience is largely affected by this. Increasing number of operators are using SIM based authentication mechanism (EAP-SIM [9]) for mobile users to ensure a seamless experience. In order to tackle the roaming issue, Wireless Broadband Alliance has recently developed the WISPr (Wireless Internet Service Provider Roaming) protocol that allows users to roam between Wi-Fi hotspots of different service providers. It uses a RADIUS server and proposes an XML based authentication protocol to seamlessly log in to hotspots without the need for the user to interact with a captive portal.

Lastly, Wi-Fi integration with DAS is not recommended in the passive DAS structure. The best option for operators is to deploy a separate network of Wi-Fi access points, thus creating another layer of Wi-Fi. This will allow best channel selection dynamically, transmit power control, and band steering.

## V. CONCLUDING REMARKS

It is expected that mobile data offloading will become a key industry segment in near future due to the unprecedented pace at which data traffic is rising on mobile networks. Wi-Fi offloading has evolved as a mature offloading solution. Most of the operators worldwide have started deploying Wi-Fi offloading solutions. However, a number of challenges exist that need to be addressed properly for creating a successful offloading mechanism. Major challenges include spatial and temporal assessment for offloading, planning and deployment issues, backhaul selection, device limitations, along with charging mechanisms. Such challenges can be properly addressed through combined effort from all the players in the value chain of mobile data offloading.